\documentclass{ws-ijmpd}
\usepackage[super,compress]{cite}
\def\sg{SGRs/AXPs}

\def\lx{L_{\rm X}}

\begin{document}

\markboth{R. V. Lobato, M. Malheiro, J. G. Coelho}
{Magnetars and White Dwarfs Pulsars}

%
\catchline{}{}{}{}{}
%

\title{Magnetars and White Dwarf Pulsars}

\author{Ronaldo V. Lobato$^{\dagger}$ and Manuel Malheiro$^{*}$}

\address{Departamento de F\'isica, Instituto Tecnol\'ogico de Aeron\'autica,\\
S\~ao J\'ose dos Campos, 12245030, S\~ao Paulo, Brazil\\
E-mail: $^{\dagger}$rvlobato@ita.br; $^{*}$malheiro@ita.br\\
http://www.ita.br/}

\author{Jaziel G. Coelho} 

\address{Divis\~ao de Astrof\'isica, Instituto Nacional de Pesquisas Espaciais - DAS/INPE/MCTI,\\
  S\~ao Jos\'e dos Campos, 12227-010, S\~ao Paulo, Brazil\\
E-mail: jaziel.coelho@inpe.br\\
  http://www.das.inpe.br/}

\maketitle

\begin{history}
\received{Day Month Year}
\revised{Day Month Year}
\end{history}

\begin{abstract}
The Anomalous X-ray Pulsars (AXPs) and Soft Gamma-ray Repeaters (SGRs) are a class of pulsars understood as neutron stars (NSs) with super strong surface magnetic fields, namely $B\gtrsim10^{14}$ G, and for that reason are known as Magnetars.  However, in the last years some SGRs/AXPs with low surface magnetic fields $B\sim(10^{12}-10^{13})$ G have been detected, challenging the Magnetar description.  Moreover, some fast and very magnetic white dwarfs (WDs) have also been observed, and at least one showed X-Ray energy emission as an ordinary pulsar. Following this fact, an alternative model based on white dwarfs pulsars has been proposed to explain this special class of pulsars. In this model, AXPs and SGRs as dense and magnetized white dwarfs can have surface magnetic field $B\sim 10^{7}-10^{10}$ G and rotate very fast with frequencies $\Omega\sim 1$ rad/s, consistent with the observed rotation periods $P\sim (2-12)$ s.
\end{abstract}


\ccode{PACS numbers: 97.60.Gb, 97.60.Jd, 04.40.Dg, 98.70.Qy}


\section{Introduction}	
The Soft Gamma-ray Repeaters and Anomalous X-ray Pulsars are a special class of pulsars, they are understood in the framework of strongly magnetized neutron star\cite{Duncan1992a, Thompson1995}, but there are alternative scenarios, in particular the white dwarf (WD) pulsar model developed by Malheiro, Rueda and Ruffini \cite{Malheiro2012}. Recently, four over a total of about 23 SGRs/AXPs presented radio-pulsed emission. These radio sources showed several properties that make them different from the others SGRs/AXPs.

As pointed by Coelho and Malheiro \cite{Coelho2014c} the large steady X-ray luminosity seen for some no-radio SGRs/AXPs, can be explained as coming from a large spin-down energy lost of a massive white dwarf with a much large magnetic dipole moment of $10^{34}\leq \mu\leq10^{36}$ emu consistent with the range observed for isolated and very magnetic WDs (see Coelho \& Malheiro 2012\cite{Coelho2012a} for discussions), indicating a different nature between these sources and the radio SGRs/AXPs. These radio SGRs/AXPs have large magnetic field and seem to be very similar to the high-B pulsars  recently founded\cite{Rea2012a}. However, even if the radio SGRs/AXPs are strong magnetized neutron stars, they are not magnetars in the sense that their steady luminosity is not originated by the magnetic energy, but from the rotational energy as rotation powered pulsars.

\section{Canonical spin-powered pulsar model}

Rotation powered pulsars (RPPs) are detected by its radio emission and their luminosity $L_X$ is considered as coming from rotation.

In this model we consider a dipole moment $\mu$ with an orientation $\alpha$ in the rotation axis that rotates with an angular frequency $\Omega$. On the surface of the star the magnetic field at the equator is $B_{\rm s}\sim\mu/R^{3}$ and at the poles $B_{\rm p}=2\mu/R^{3}$, where $R$ is the star radius. The rotational energy is given by \cite{Shapiro2008},
\begin{equation}
E_{\textrm{rot}}=\frac{1}{2}I\Omega^2,
\end{equation}
where $I$ is the moment of inertia. Thus, the loss of rotational energy of the pulsar is:
\begin{eqnarray}\label{erot}
\dot{E}_{\rm{rot}}=I\Omega\dot\Omega+\frac{1}{2}\dot{I}\Omega^2\approx I\Omega\dot\Omega.
\end{eqnarray}
Such configuration has a time-varying magnetic dipole moment, in an infinity referential, and radiates energy with a rate
\begin{eqnarray}\label{edip}
\dot{E}_{\rm{dip}}=\frac{2\mu^2\Omega^4\sin^2\alpha}{3c^3},
\end{eqnarray}
where $c$ is the speed of light. Combining (\ref{erot}) and (\ref{edip}) we obtain the surface magnetic field at the equator as \cite{Ferrari1969}:
\begin{eqnarray}\label{campo}
B_{\rm s}=B_{\rm p}/2=\left(\frac{3c^3I}{8\pi^2R^6}P\dot{P}\right)^{1/2},
\end{eqnarray}
with period $P=2\pi/\Omega$, and $\dot{P}=dP/dt$.
We see that the value of the magnetic field is inferred from $\sqrt{P\dot{P}}$ that are measures obtained by astronomical observations, and two parameters, the moment of inertia $I$ and the star radius $R$ that depend how the mass-energy density is distributed inside the star, are obtained solving the Tolman-Oppenheimer-Volkoff equation for a specific equation of state chosen, and of course, are quite different for Neutron Stars or White Dwarfs.

\section{The magnetar model}
In the magnetar model, the SGRs/AXPs are considered as neutron stars with large magnetic fields, higher than the value of critical field from the quantum electrodynamics (QED) $B_q\equiv m^2c^3/\hbar e=4.4\times10^{13}$ G ($m$ is the electron's mass, $c$ the speed of light and $\hbar$ the reduced Planck's constant and $e$ the charge) and 100-1000 times than the magnetic field of normal pulsars observed.  The observed X-ray luminosity $L_X$ is determined by complex structures in its magnetosphere (see \cite{Duncan1992a,Thompson1995}), considering that luminosity is powered by the energy of magnetic field.

The recent discovery of radio-pulsed emission in four of this class of sources, where the spin-down rotational energy lost $\dot{E}_{\rm rot}$ is larger than the X-ray luminosity $L_X$ as we see in the Figure~\eqref{flxerotns} opens the question of the nature of these radio sources in comparison to the other SGRs/AXPs (we have also recently studied this point in Refs.~\cite{Coelho2013b, Lobato2015,Lobato2015a}, where the models of radio emission were investigated).

\subsection{Difficulties with the magnetar model}
According the magnetar model the high X-ray luminosity from this sources is powered by the energy stored in their strong magnetic fields~\cite{Mereghetti2013b, Turolla2015a}. Although the interpretation as neutron stars is the one most successfully, here are some aspects about \sg\ not very well understood in this interpretation:
\begin{itemize}
\item RPPs are assumed born with a short period $\sim$1ms resulting in large magnetic fields by dynamo process \cite{Tong2011}, the \sg\ presents longs period, such that generation of high magnetic fields by this mechanisms is not well understood.
\item the large spin-down rates in comparison with the typical pulsar implies, for example, that the source SGR 0418+5729 has a characteristic age $\tau=P/2\dot{P}=2.4\times10^{7}$ years. Thus, it is difficult to understand, how being  much older than ordinary pulsars, it still fits in the magnetar model where the sources are seen as young pulsars. 
  \end{itemize}
In fact, some \sg\ with low magnetic fields and smaller spin-down rates were observed \cite{Rea2010}. In this observation we highlight SGR 0418+5729, with period $P=9.08$s, spin-down $\dot{P}<6.0\times10^{-15}$s/s and X-ray luminosity $L_{\rm X}<6.2\times10^{31}$erg/s, considering this source a neutron star, the loss of energy associated with injection of rotational energy in the pulsars magnetosphere does not explain the luminosity of this star, i.e., $\dot{E}_{\rm rot}^{\rm NS}<L_{\rm X}$. Thus,  excludes the possibility to be a RPP, where $\dot{E}_{\rm rot}^{\rm NS}>L_{\rm X}$. Recent observations of Fermi satellite \cite{Abdo2010, Tong2010, Tong2011a} does not find evidences of $\gamma$ radiation. Furthermore, were detected 4 \sg\ that emit in wavelength of radio, while for the others not. These observations carried some uncertainties about the nature of \sg\ and opens the possibility of consider a different nature for these sources, as white dwarf pulsars as proposed by Malheiro, Rueda and Ruffini\cite{Malheiro2012}. 

\section{SGRs/AXPs as white dwarf pulsars}
Rapidly, massive white dwarfs and with observed high magnetic field $10^{6}-10^{9}$G were detected in the recent years \cite{Castanheira2013} and share some similarities with \sg. After these facts, Malheiro and collaborators \cite{Malheiro2012} following two works from Morini\cite{Morini1988} and Paczynski\cite{Paczynski1990} developed an alternative model  explaining \sg\ as white dwarfs. As was investigated by Usov in 1989 \cite{Usov1988}, the process of release energy in a massive, magnetic white dwarf can be explained in terms of a canonical spin-powered pulsars model, because in several aspects they are similar.\\
For example, if we consider a star with $M=1.4{M_{\odot}}$ and $R=10^{6}\ \rm{cm}$ (a neutron star), from  (\ref{campo}) the magnetic field at poles is:
\begin{eqnarray}
B_{\rm p}^{\rm NS}=3.2\times 10^{19}(P\dot{P})^{1/2}{\rm G},
\end{eqnarray}
In the case of a white dwarf with $M=1.4M_{\odot}$ and $R=3\times 10^8\ \rm{cm}$, there is a new scale for the magnetic field at poles:
\begin{eqnarray}
B_{\rm p}^{\rm WD}=4.21\times 10^{14}(P\dot{P})^{1/2}{\rm G},
\end{eqnarray}
and to the dipole moment $\mu$ , responsible by the dipole radiation emitted,   which is a factor $10^{3}$ times larger than for neutron stars. That is the factor observed between the X-Ray luminosity of \sg\ as white dwarfs and slowly pulsars like the XDINS \cite{Kaplan2009}, and high magnetic pulsars \cite{Gavriil2008}: basically all these sources present the same period $P\sim1-10$s that \sg. When \sg\ are white dwarfs there are new values for the mass density, moment of inertia, dipole moment and rotation energy. It explains a large class of problems related in considering SGRs/AXPs as neutron stars. Furthermore, we cannot ignore the recent astronomical observations of old SGRs (characteristic age $\sim(10^6-10^7)$ Myr) with low surface magnetic field, as well as the four SGRs/AXPs showing radio emission recently discussed by Coelho and Malheiro\cite{Coelho2013a, Coelho2013}.
In the table \eqref{kstars} we summarized the properties among SGR 0418+5729, Swift J1822.3-1606 \cite{Rea2013} (both SGRs with low-B), AE Aquarii~\cite{Terada2008b} (the first white dwarf pulsar), RXJ 0648.0-4418 (rapid WD that emit in X-ray), and the candidate EUVE J0317-855 \cite{Ferrario1997}.

\begin{table}[ph]
\tbl{Comparison between the observational and deduce proprieties in the dipole model, where the \sg\ were consider as white dwarfs: the quantities $P$, $\dot{P}$ and $L_{\rm X}$ were obtained in \cite{Olausen2014}.}
{\begin{tabular}{@{}lcccccc@{}} \toprule
Stars & $P$(s) & $\dot{P}$($10^{-14}$ s/s) & $\tau$($10^6$ years) & $L_{\textrm{X}}$(erg/s) & $B_{\textrm{\rm e}}^{\rm WD}$(G) & $\mu_{\textrm{p}}^{\rm WD}$ (emu) \\ \colrule
SGR 0418    & 9.08  & $<0.6$   & 24   & $\sim6.2\times10^{31}$ & $<9.83\times10^{7}$   & $2.65\times10^{33}$     \\
Swift J1822 & 8.44  & 8.3    & 1.6  & $\sim4.2\times10^{32}$ & $3.52\times10^{8}$    & $0.95\times10^{34}$     \\
AE Aquarii  & 33.08 & 5.64   & 9.3  & $\sim10^{31}$          & $\sim5\times10^{7}$   & $\sim1.35\times10^{33}$ \\
RX J0648    & 13.2  & 0.6    & 0.23 & $\sim10^{32}$          & $0.1\times10^{9}$     & $3.48\times10^{34}$     \\
EUVE J0317  & 725   & -      & -    & -                     & $\sim4.5\times10^{8}$ & $1.22\times10^{34}$     \\ \botrule
\end{tabular} \label{kstars}}
\end{table}

In the description of \sg\ as white dwarfs some problems inherent to magnetar models are solved as consequence of the new scales:
\begin{itemize}
\item the existence of stable white dwarfs could explain the band of period $2\lesssim P\lesssim12$s observed in \sg. In particular the fact that we do not observe \sg\ with $P<2$s is due the white dwarf structure. In \cite{Boshkayev2013a} self-consistent computations shown that the minimum period to white dwarfs are $\sim0.3$, $0.5$, $0.7$ and $2.2$s to star made of $^{4}$He, $^{12}$C, $^{16}$O, $^{56}$Fe respectively.
\item the energy balance is solved, since the X-luminosity observed is less than the rotational energy of a white dwarf, $L_{\rm X}<\dot{E}_{\rm rot}$, because $\dot{E}_{\rm rot}$ is $10^{5}$ bigger than the value to a neutron star. \sg\ powered by rotational energy are completely analogous to RPPs of neutron star.
\item largest luminosity $L_{\rm X}\sim10^{35}$erg/s to slowly pulsar ($\Omega\sim1$Hz) is understood as consequence of the white darfs' radius that produces a large dipole moment. Some works shown the possibility of pulsed radiation in white dwarfs. \sg\ like white dwarfs have large magnetic fields and rotate very fast, near the Kepler frequency, and they could produce large potential difference on its surface able to produce X and $\gamma$ radiation \cite{Kashiyama2013}. Bursts in this sources, that is common, would be a consequence of their angular velocity and change in the moment of inertia. In this situation there are blunt changes in gravitational energy \cite{Malheiro2012}.
  \item \sg\ has a low population in contrast with more than 2000 RPPs founded, this is justified by the motivation that white dwarfs with high magnetic field an rapidly are rarely formed \cite{Ilkov2012}. Furthermore following the observations, white dwarfs with high magnetic fields are only 10\% of magnetic stars.

\end{itemize}

\subsection{Efficiency}
\begin{figure}[h!]
\resizebox{\hsize}{!}{\includegraphics{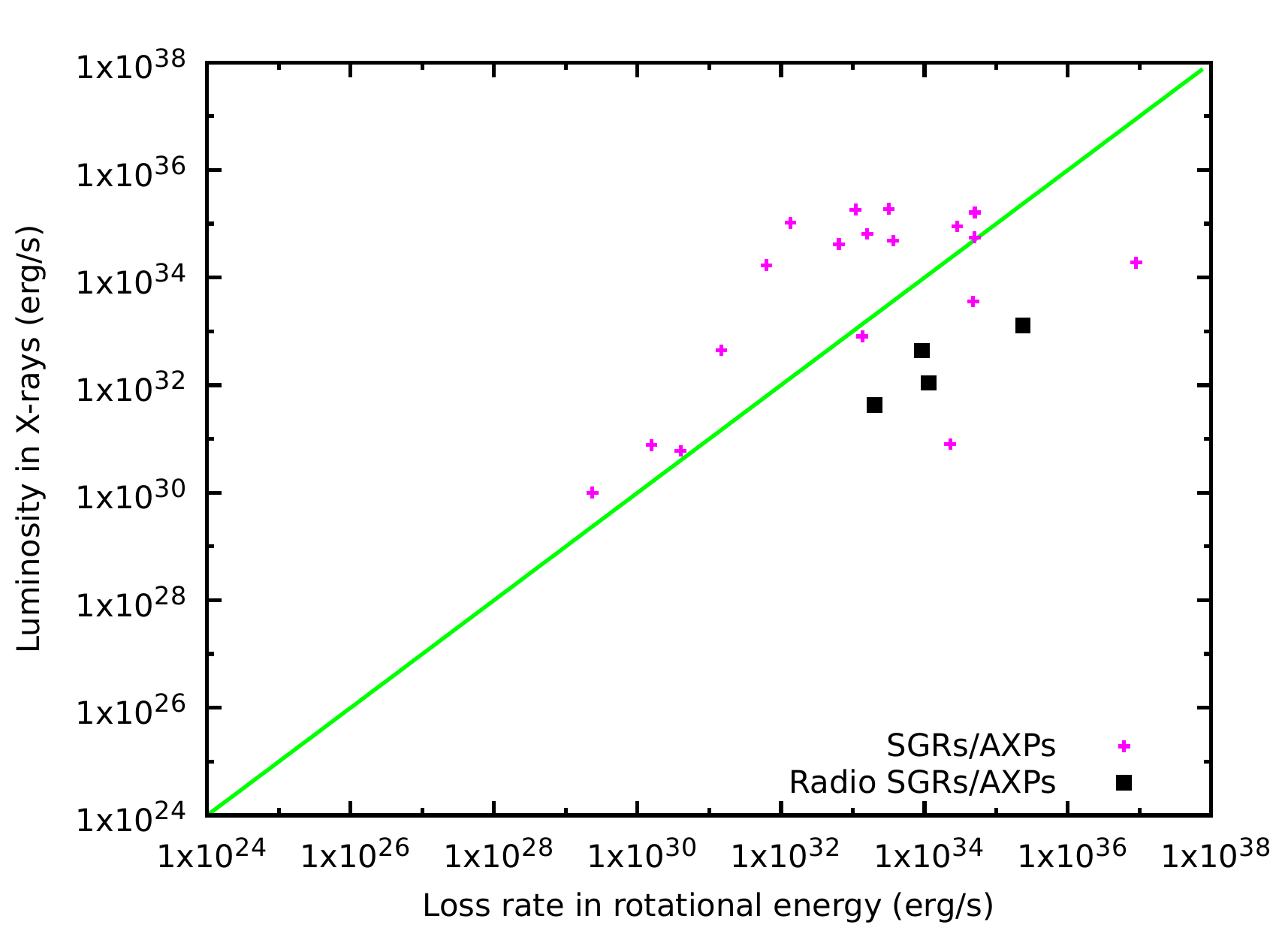}}
\caption{Four SGRs/AXPs in comparison with all no-radio SGRs/AXPs as neutron stars. These radio sources have large magnetic fields and are very similar to high-$B$ pulsars. Even this sources are strong magnetized neutron stars, they are not magnetar, because their steady luminosity can be explained by loss of rotational energy.}
\label{flxerotns}
\end{figure}
\sg\ when understood as neutron stars typically have a larger luminosity in X-ray that can not be explained by their spin-down luminosity, i.e., $\lx\gg\dot{E}_{\rm rot}$. However, in four of them that is not true and they present radio emission. These radio \sg\ showed several properties different from the others \sg. As was discussed, in this four sources the X-ray efficiency $\eta_{\rm X}=L_{\rm X}/{\dot{E}_{\rm rot}}$ seems to be small comparing with \sg\ as white dwarfs\cite{Malheiro2012, Coelho2013}. However, when these four sources are understood as neutron stars pulsars the efficiency is $\eta_{\rm X}\sim$(0.2-0.1), much larger than the values of ordinary pulsars where $\eta_{\rm X}\sim$ ($10^{-3}$ to $10^{-4}$), but still smaller than one: for this reason Coelho and Malheiro \cite{Coelho2013} suggested that these four star are very efficient  neutron stars RPPs in contrast to the others that are white dwarfs RPPs. As we can see in the figure \eqref{flxerotns} this four sources presents a loss of rotational energy larger than the X-ray luminosity. They are very similar to the high-$B$ pulsars recently found\cite{Olausen2010, Olausen2013a} which have their luminosity powered by rotational energy.

\subsection{Difficulties with the white dwarf pulsar model}
Although SGRs/AXPs as white dwarfs pulsars are well described, there are some difficulties aspects in the phenomenology: 
\begin{itemize}
\item Some SGRs/AXPs are associate with supernova. Usually, this association is an evidence of SGRs/AXPs to be neutron stars; and also some are associate with new born stars' clusters.
\item AXPs show up variation in its luminosity and some transients seems to contradict the rotation powered hypotheses.
\end{itemize}
SGRs/AXPs as white dwarfs pulsars in comparison with magnetar have larger radius. However is difficult to estimate the radius due to the distance, since the majority of SGRs/AXPs are in $d\gtrsim 2$ kpc. The knowledge about their nature has an uncertainty until we have better measures for the radii of these sources.

\section{Conclusions}
Some SGRs/AXPs within the description of white dwarfs pulsars can be well understood. In the white dwarfs pulsars scenario we solve the majority of problems concerning the magnetar model, in particular the ultra high magnetic fields. Super-massive, fast and highly magnetized white dwarfs have already been found and, this number is increasing every year, opening a new research area. The future will solve the correct nature of SGRs/AXPs: Magnetars or White Dwarf pulsars. 

\section*{Acknowledgments}

This work was financially supported by CAPES (Coordena\c c\~ao de Aperfei\c coamento de Pessoal de N\'ivel Superior), CNPq (Conselho Nacional de Desenvolvimento Cient\'ifico e
Tecnol\'ogico) and FAPESP thematic project 2015/26258-4. J.G.C. acknowledges the support of FAPESP (2013/15088-0 and 2013/26258-4). Acknowledgements 
also to organizing committee for the hospitality in the III Amazonian Symposium on Physics and V NRHEP Network Meeting: Celebrating  100 years of General Relativity.

\bibliographystyle{ws-procs975x65}
\bibliography{library}

\end{document}